\documentclass{article}
\usepackage{waspaa21,amssymb,amsmath,times,url}
\usepackage[dvipdfmx]{graphicx}
\usepackage{color}

\usepackage{bm,cite,subfig}
\usepackage{algorithm,algorithmic}

\title{MeshRIR: a dataset of room impulse responses on meshed grid points for evaluating sound field analysis and synthesis methods}

\name{Shoichi Koyama,
      Tomoya Nishida,
      Keisuke Kimura,
      Takumi Abe,
      Natsuki Ueno, 
      and Jesper Brunnstr\"{o}m}
\address{The University of Tokyo, 7-3-1 Hongo, Bunkyo-ku, Tokyo 113-8656, Japan\\ koyama.shoichi@ieee.org
}

\begin{document}

\ninept
\maketitle

\begin{sloppy}

\begin{abstract}
A new impulse response (IR) dataset called ``MeshRIR'' is introduced. Currently available datasets usually include IRs at an array of microphones from several source positions under various room conditions, which are basically designed for evaluating speech enhancement and distant speech recognition methods. On the other hand, methods of estimating or controlling spatial sound fields have been extensively investigated in recent years; however, the current IR datasets are not applicable to validating and comparing these methods because of the low spatial resolution of measurement points. MeshRIR consists of IRs measured at positions obtained by finely discretizing a spatial region. Two subdatasets are currently available: one consists of IRs in a three-dimensional cuboidal region from a single source, and the other consists of IRs in a two-dimensional square region from an array of 32 sources. Therefore, MeshRIR is suitable for evaluating sound field analysis and synthesis methods. This dataset is freely available at \url{https://sh01k.github.io/MeshRIR/} with some codes of sample applications.
\end{abstract}


\section{Introduction}
\label{sec:intro}

Acoustic room impulse responses (IRs) contain characteristics of sound propagation between a source and a receiver as a linear time-invariant system. Since numerical simulations of acoustic waves in a three-dimensional (3D) space require a high computational load, the IRs measured in practical environments are useful for evaluating the performance of acoustic signal processing techniques, such as speech enhancement, dereverberation, source localization, source separation, and speech recognition, as well as auralization. Therefore, many IR datasets have been developed by various research communities~\cite{Nakamura:ASJ1999,Wen:IWAENC2006,Jeub:DSP2009,Stewart:ICASSP2010,Shelley:AESconv2010,Kinoshita:WASPAA2013,Eaton:WASPAA2015,Szoke:IEEE_JSTSP2019}. 

Currently available public IR datasets usually contain IRs of several loudspeaker and microphone placements (or devices) under several reverberation conditions. For example, the Real World Computing Partnership Sound Scene Database (RWCP-SSD)~\cite{Nakamura:ASJ1999} includes IRs measured by circular and linear microphone arrays from a loudspeaker at several positions in nine different rooms. The Aachen Impulse Response (AIR) dataset~\cite{Jeub:DSP2009} includes IRs measured by a dummy head and a mobile phone at several positions in six rooms. The BUT Speech@FIT Reverb Database (BUT ReverbDB)~\cite{Szoke:IEEE_JSTSP2019} consists of IRs at 31 microphone positions from 2--11 loudspeaker positions in nine rooms. Recently, a dataset of IRs from loudspeakers at cube-shaped grid to linear microphone arrays has been developed~\cite{Cmejla:EUSIPCO2020}. These datasets are designed mainly for evaluating speech enhancement and distant speech recognition methods. 

On the other hand, sound field analysis and synthesis methods have been extensively investigated in recent years; these methods include the estimation and visualization of acoustic fields~\cite{Chardon:JASA2012,Grande:JASA2016,Koyama:IEEE_J_JSTSP2019}, sound field recording for spatial audio applications~\cite{Samarasinghe:IEEE_ACM_J_ASLP2014,Ueno:IEEE_SPL2018,Ueno:IEEE_J_SP_2021}, the interpolation of acoustic transfer functions~\cite{Mignot:IEEE_J_ASLP2013}, and sound field synthesis and control with multiple loudspeakers~\cite{Spors:AES124conv,Poletti:J_AES_2005,Ueno:IEEE_ACM_J_ASLP2019,Koyama:IEEE_ACM_J_ASLP2020}. To evaluate these methods, characteristics of sound propagation inside a spatial region must be known. Therefore, in many of these studies, experimental evaluations are limited to numerical simulations. To validate and compare the sound field analysis and synthesis methods in more realistic scenarios, a publicly available IR dataset at finely meshed grid points inside a spatial region is necessary. The IR dataset developed by Steward and Sandler (C4DM RIR dataset)~\cite{Stewart:ICASSP2010} includes IRs in a two-dimensional (2D) rectangular region in three different rooms, but the intervals between measurement points ($0.5$--$1.0~\mathrm{m}$) are not sufficiently small for this purpose.

We introduce a new IR dataset called ``MeshRIR'' designed for the evaluation of sound field analysis and synthesis methods, which has the following features: 
\begin{itemize}
 \item Measurement points are set at finely meshed grid points inside a spatial region. The intervals between the measurement points are $0.05~\mathrm{m}$.
 \item Two types of measurement region are currently available: 3D cuboidal and 2D square, whose dimensions are $1.0~\mathrm{m} \times 1.0~\mathrm{m} \times 0.4~\mathrm{m}$ and $1.0~\mathrm{m} \times 1.0~\mathrm{m}$, respectively. 
 \item IRs from a single source position to a 3D cuboidal region and from an array of 32 sources to a 2D square region are included. 
 \item Several codes for sample applications are also provided, including sound field reconstruction for estimating the pressure distribution from a discrete set of measurements and sound field control for synthesizing a desired field inside a target region.
\end{itemize}
The dataset is publicly available at \url{https://sh01k.github.io/MeshRIR/} under a nonrestrictive license. 

%Impulse response dataset
%\begin{itemize}
% \item Real World Computing Partnership Sound Scene Database (RWCP-SSD)~\cite{Nakamura:ASJ1999}
% \item Multichannel Acoustic Reverberation Database at York (MARDY)~\cite{Wen:IWAENC2006}
% \item Aachen Impulse Response (AIR)~\cite{Jeub:DSP2009}
% \item C4DM RIR Dataset~\cite{Stewart:ICASSP2010}
% \item OpenAIR~\cite{Shelley:AESconv2010}
% \item REVERB Challenge~\cite{Kinoshita:WASPAA2013}
% \item ACE Challenge~\cite{Eaton:WASPAA2015}
% \item Brno University of Technology Speech@FIT Reverberation Database (BUT ReverbDB)~\cite{Szoke:IEEE_JSTSP2019}
%\end{itemize}

%\newpage

\section{Details of dataset}

\begin{table*}[t]
 \centering
 \caption{Detailed measurement conditions of subdatasets \textbf{S1-M3969} and \textbf{S32-M441}.}
 \label{tbl:info}
 \begin{tabular}{l|c|c}
 \hline
  & \textbf{S1-M3969} & \textbf{S32-M441} \\ \hline \hline
 Sampling rate & \multicolumn{2}{|c}{$48000~\mathrm{Hz}$} \\ \hline
 IR length & \multicolumn{2}{|c}{$32768~\mathrm{samples}$} \\ \hline
 Room dimensions & \multicolumn{2}{|c}{$7.0~\mathrm{m} \times 6.4~\mathrm{m} \times 2.7~\mathrm{m}$} \\ \hline
 Number of source positions & $1$ & $32$\\ \hline
 Measurement region & 3D cuboidal: $1.0~\mathrm{m} \times 1.0~\mathrm{m} \times 0.4~\mathrm{m}$ & 2D square: $1.0~\mathrm{m} \times 1.0~\mathrm{m}$ \\ \hline
 Intervals of microphone positions & \multicolumn{2}{|c}{$0.05~\mathrm{m}$} \\ \hline
 Number of microphone positions & $21 \times 21 \times 9~(= 3969)~\mathrm{points}$ & $21 \times 21~(= 441)~\mathrm{points}$ \\ \hline
 Reverberation time ($T_{60}$) & $0.38~\mathrm{s}$ & $0.19~\mathrm{s}$ \\ \hline
 Average temperature & $26.3^\circ$C & $17.1^\circ$C \\ \hline
 \end{tabular}
\vspace{-10pt}
\end{table*}

MeshRIR consists of two subdatasets. One consists of IRs inside a 3D cuboidal region from a single source position (\textbf{S1-M3969}). The other consists of IRs inside a 2D square region from 32 source positions (\textbf{S32-M441}). The measurement conditions are summarized in Table~\ref{tbl:info}. In both subdatasets, the measurement points were determined by discretizing the measurement region every $0.05~\mathrm{m}$. All the IRs were measured in a room with approximate dimensions of $7.0~\mathrm{m} \times 6.4~\mathrm{m} \times 2.7~\mathrm{m}$. The sampling frequency was $48000~\mathrm{Hz}$ and the IR length was $32768~\mathrm{samples}$.

\subsection{Equipment and conditions for IR measurement}

To achieve a low measurement error, the excitation signal for measuring IRs should have a sufficiently high output level at all target frequencies while keeping nonlinear errors of the measurement system low. There are two major categories of excitation signals for IR measurement: maximum length sequence (MLS) and swept-sine signal~\cite{ISO18233:2006,Stan:JAES2002}. The MLS is a periodic pseudorandom signal whose statistical properties are similar to those of white Gaussian noise~\cite{Schroeder:JASA1979,Rife:JAES1989,Dunn:JAES1993}. The swept-sine signal is chirp signal whose frequency increases or decreases with time, and this signal is also called the time-stretched pulse (TSP)~\cite{Berkhout:JASA1980,Muller:JAES2001}. There are several variations in the change in frequency. The increase or decrease in the instantaneous frequency of the linear swept-sine signal is linear~\cite{Aoshima:JASA1981,Suzuki:JASA1995}. On the other hand, the instantaneous frequency of the exponential swept-sine signal exponentially increases, as its name suggests~\cite{Farina:AESconv2000,Moriya:AST_AL2005,Farina:AESconv2007}. The exponential swept-sine signal is sometimes called logarithmic swept-sine signal since its group delay logarithmically changes.

The difference between the excitation signals appears in their power spectrum and nonlinear error. Whereas the power spectra of the MLS and linear swept-sine signal are flat as in that of white noise, that of the exponential swept-sine signal linearly decreases with increasing frequency as in that of pink noise. Thus, the SNR of the excitation signals with a flat power spectrum is proportional to the noise power spectrum. The SNR of the exponential swept-sine signal can be relatively higher at low frequencies than at high frequencies. The nonlinear error of the MLS arises as distortion artifacts almost uniformly distributed in the time-frequency domain. When using the linear swept-sine signal with increasing frequency, major parts of harmonic distortions appear before the initial linear response as pre-echos. 
%because the time-frequency spectra of the harmonic distortions have steeper slope than that of the linear response, and they are shifted before that of the linear response by convoluting with its inverse signal. 
The same thing happens when using the exponential swept-sine signal, but the harmonic distortion of each order is separately concentrated on specific time points. 
%Thus, the harmonic distortions can be separated from the linear response in the time domain. 
In the linear swept-sine signal with decreasing frequency, the harmonic distortions are overlapped with the linear response.
%, but they do not appear before the initial linear response. 

The selected excitation signal for our purpose was the linear swept-sine signal with decreasing frequency. Since we carefully adjusted the output level of the loudspeakers, the nonlinear effects will be negligible. 
%This is because it is difficult to remove nonlinear effects, i.e., pre-echos, from thousands of measured IRs by postprocessing. Therefore, we carefully adjusted the output level. 
We also designed the linear swept-sine signals in the frequency domain to ensure that all the frequency components up to the Nyquist frequency are equally included without discontinuities at the beginning and end of the signals~\cite{Suzuki:JASA1995,Muller:JAES2001}. The IRs were reconstructed by the circular convolution of inverse signals after synchronous averaging three times . 

To measure the IRs on the grid points, we used a Cartesian robot with an omnidirectional microphone (Primo, EM272J) to control the measurement position in three dimensions (see Fig.~\ref{fig:system}). The signal input and output were controlled by a PC with a Dante interface. We also monitored the room temperature as each IR was measured. The file formats of IR data are MAT for MATLAB and NPY for Python. All the additional data are provided as a JSON file.

\begin{figure}[!t]
  \centering
  \includegraphics[width=0.8\columnwidth,clip]{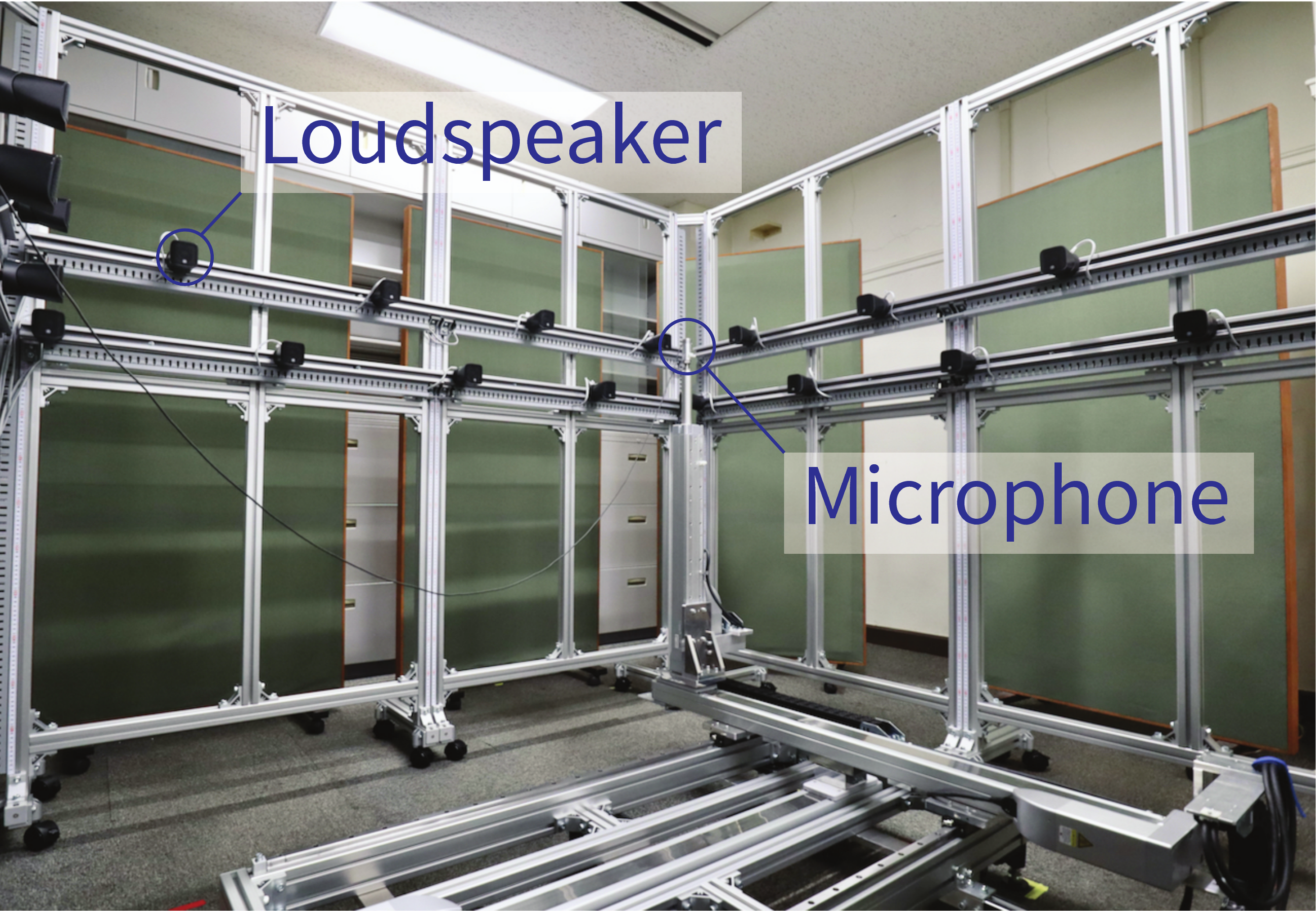}
  \caption{IR measurement system for \textbf{S32-M441}.}
  \label{fig:system}
 \vspace{-5pt}
%\end{figure}
%\begin{figure}[!t]
  \centering
  \subfloat[\textbf{S1-M3969}]{\includegraphics[width=1.6in,clip]{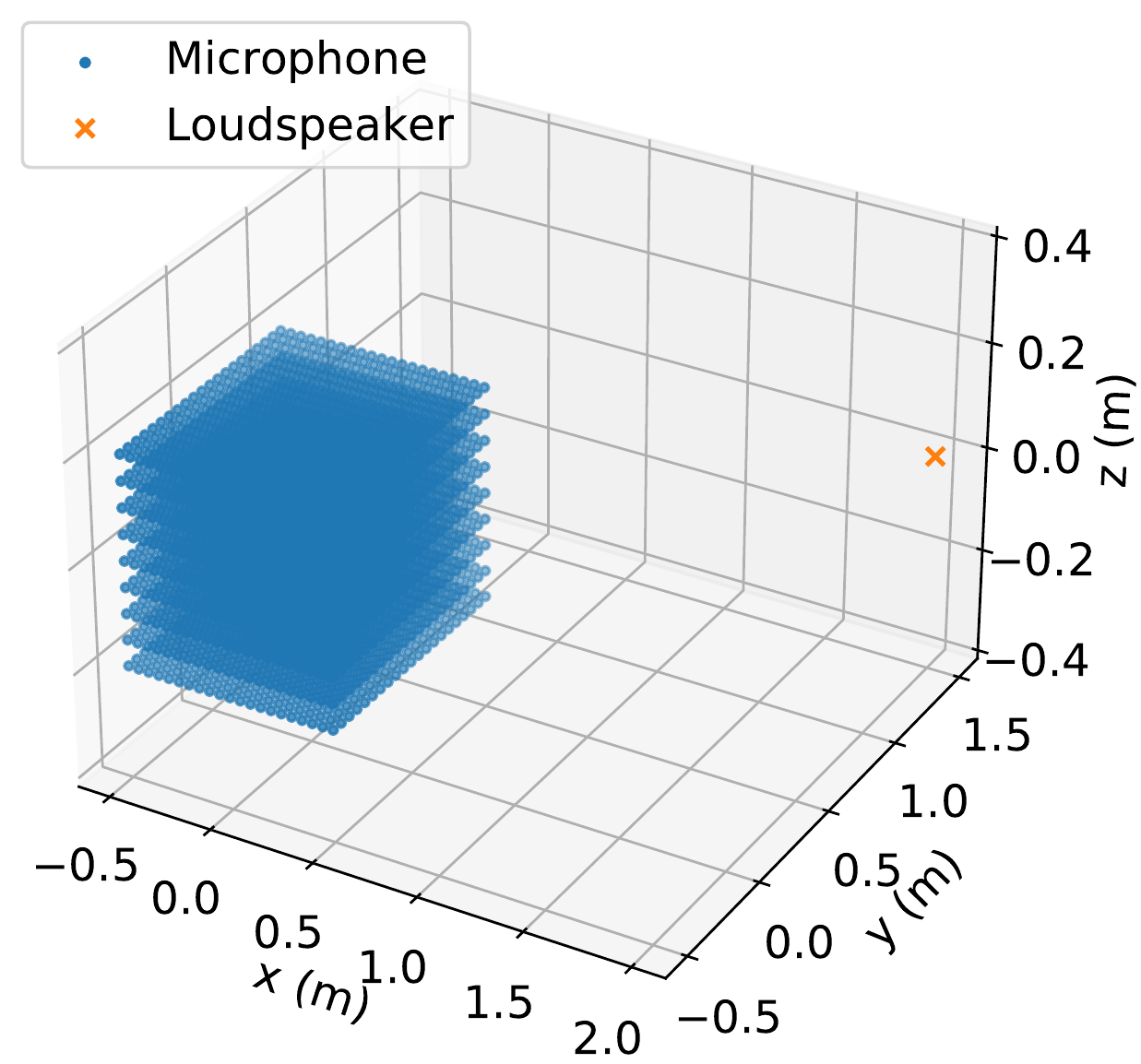}} \ 
  \subfloat[\textbf{S32-M441}]{\includegraphics[width=1.6in,clip]{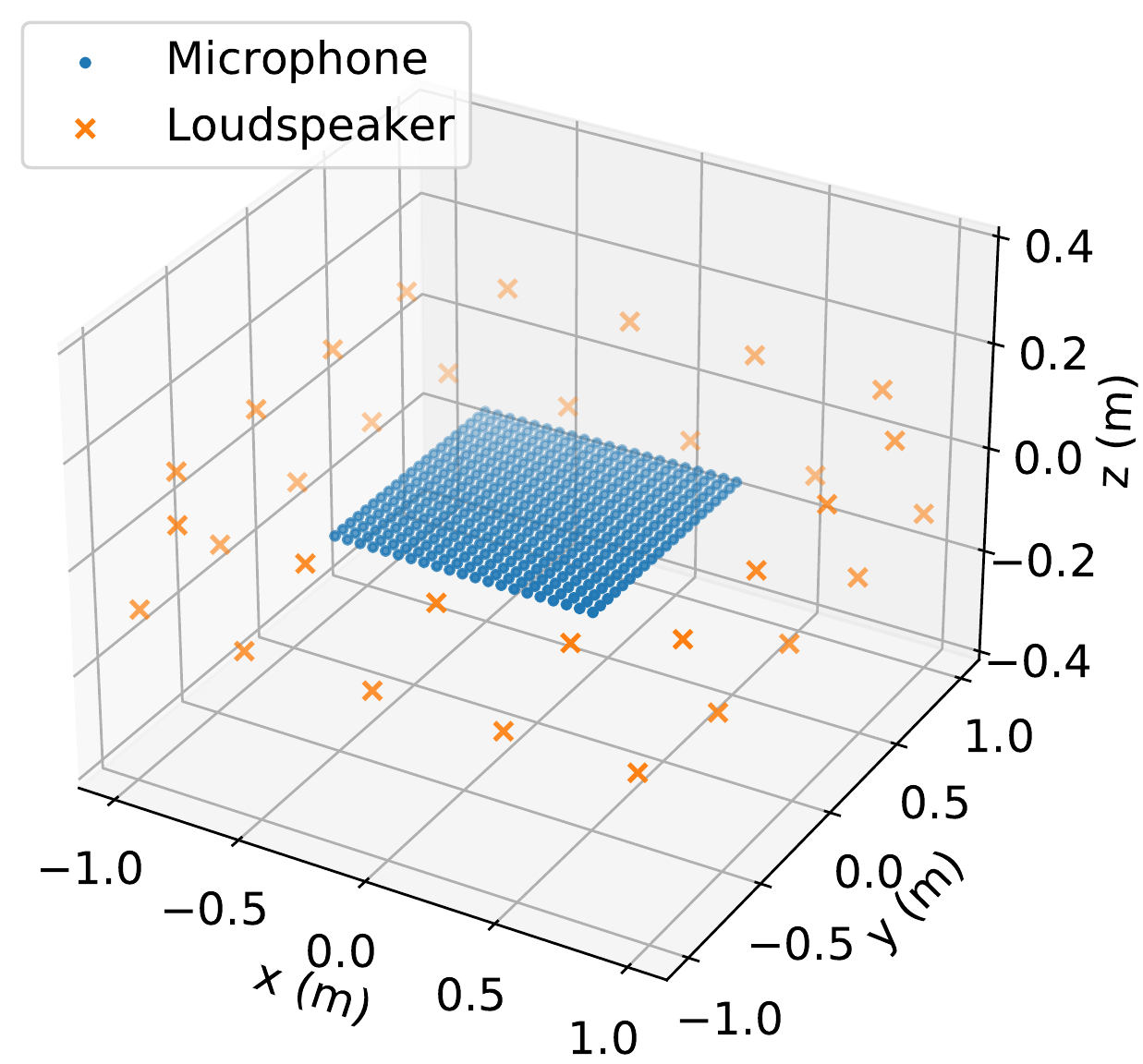}}
  \caption{Positions of loudspeakers and microphones.}
  \label{fig:pos}
\vspace{-5pt}
\end{figure}

\subsection{S1-M3969}

In \textbf{S1-M3969}, the IRs from a single source at the fixed position to the 3D cuboidal region with dimensions of $1.0~\mathrm{m} \times 1.0~\mathrm{m} \times 0.04~\mathrm{m}$ were included. The coordinate origin was set at the center of the measurement region, and the source position was $(2.0, 1.5, 0.0)~\mathrm{m}$. The measurement points were obtained by discretizing the cuboidal region at intervals of $0.05~\mathrm{m}$, so the number of microphone positions was $21 \times 21 \times 9$ ($=3969$)~$\mathrm{points}$. Fig.~\ref{fig:pos}(a) shows the loudspeaker and microphone positions. The source was an ordinary closed loudspeaker (DIATONE, DS-7). The reverberation time ($T_{60}$) averaged for randomly chosen $100$ IRs was $0.38~\mathrm{s}$. This sub-dataset will be useful for evaluating 3D sound field analysis methods, e.g., the interpolation of pressure fields and acoustic transfer functions. 

\subsection{S32-M441}

The subdataset \textbf{S32-M441} was the set of IRs from an array of sources to a planar measurement region. As shown in Fig.~\ref{fig:pos}(b), the loudspeakers were regularly placed along the borders of two squares with dimensions of $2.0~\mathrm{m} \times 2.0~\mathrm{m}$ at heights of $z=-0.2~\mathrm{m}$ and $0.2~\mathrm{m}$. Small closed loudspeakers (YAMAHA, VXS1MLB) were used. The measurement region was a square with dimensions of $1.0~\mathrm{m} \times 1.0~\mathrm{m}$. Again, the measurement region was discretized at intervals of $0.05~\mathrm{m}$, so the number of microphone positions was $21 \times 21$ ($=441$)~$\mathrm{points}$. Several acoustic absorption panels were placed. The reverberation time ($T_{60}$) averaged for randomly chosen IRs of $10$ sources and $100$ microphones was $0.19~\mathrm{s}$. This subdataset is useful for evaluating sound field synthesis and control methods.

\section{Evaluation and sample applications}

We show some evaluation results of the dataset. Two sample applications of the dataset, which are sound field reconstruction and control, are also shown.

\subsection{Evaluation of IRs}

Since the dataset includes IRs on finely meshed grid points, the spatial sound field can be visualized. We extracted IRs at $z=0$ of \textbf{S1-M3969}, which is the same as the height of the source, and filtered each IR using a low-pass filter, whose cutoff frequency was $600~\mathrm{Hz}$, for visibility. 

Fig.~\ref{fig:dist_xy} shows the instantaneous pressure distributions at $t=8480$, $8520$, $8560$, and $8600~\mathrm{samples}$. The direct wave propagating on the $x$-$y$-plane is visualized. The IRs with respect to time along the $y$-axis at $(x,z)=(0,0)~\mathrm{m}$ are plotted in Fig.~\ref{fig:dist_yt}. The reflected waves following the direct wave can be seen. 

\begin{figure}[!t]
  \centering
  \subfloat[$t=8480~\mathrm{sample}$]{\includegraphics[width=1.7in,clip]{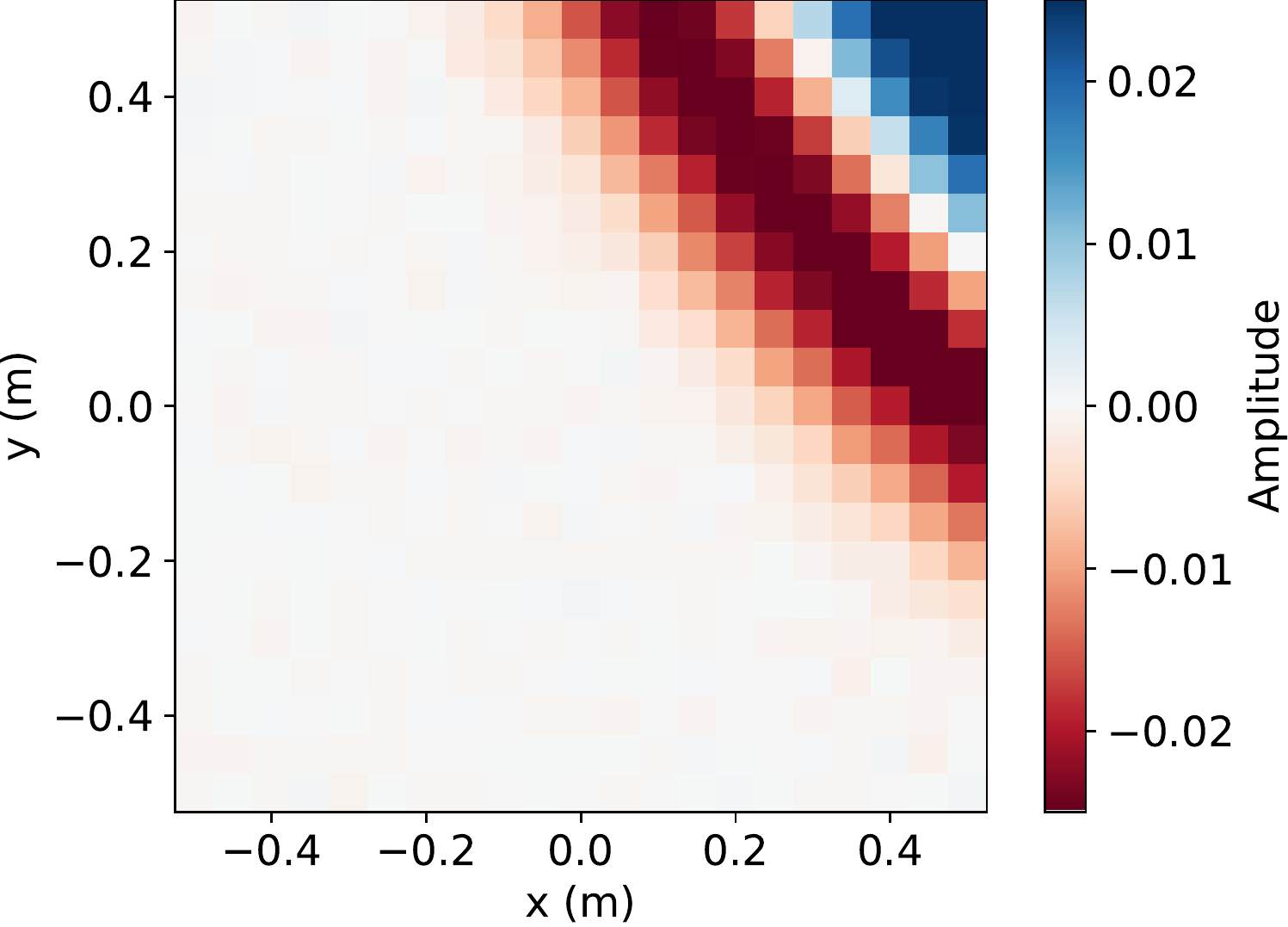}}
  \subfloat[$t=8520~\mathrm{sample}$]{\includegraphics[width=1.7in,clip]{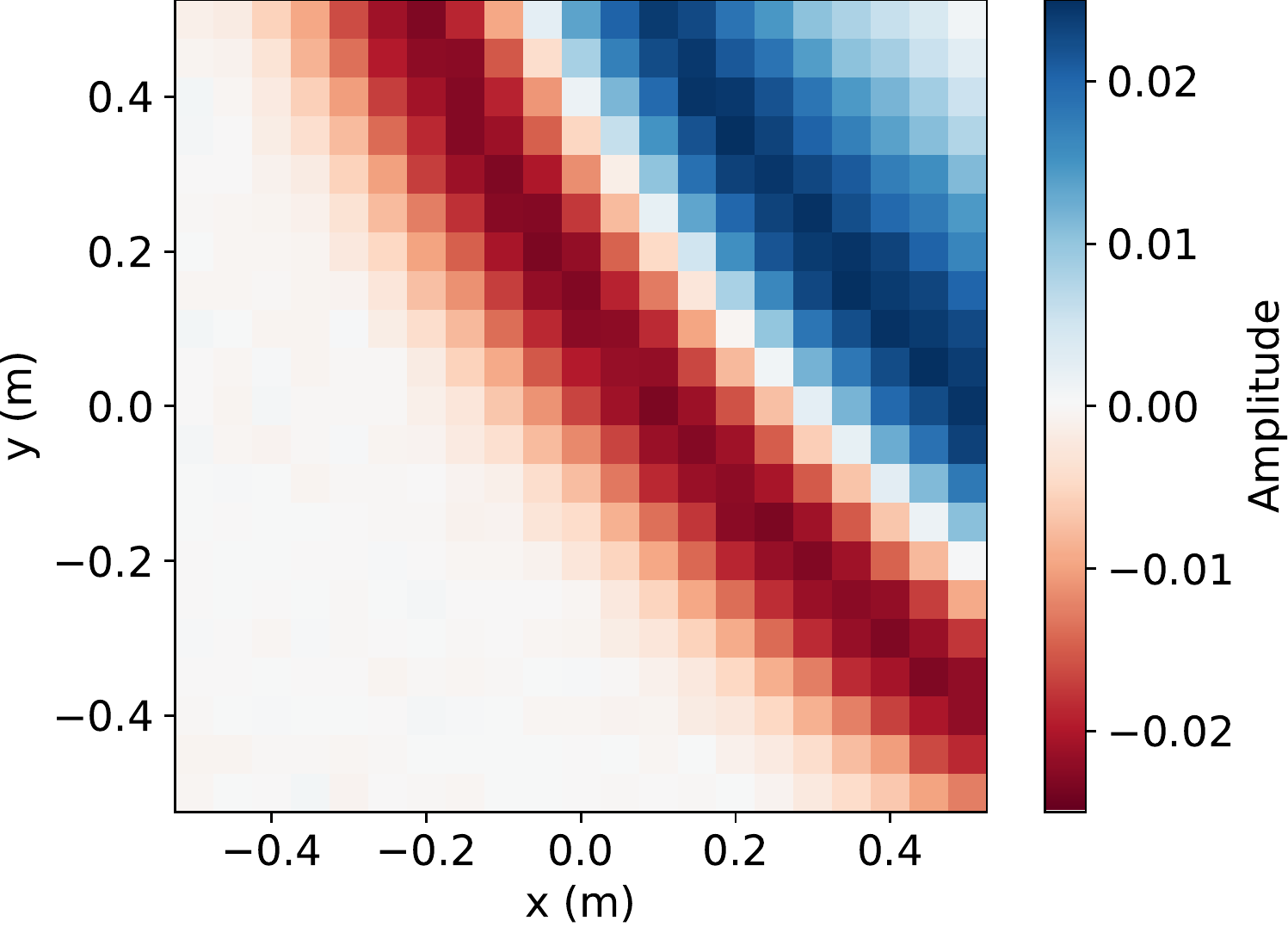}}\\
  \subfloat[$t=8560~\mathrm{sample}$]{\includegraphics[width=1.7in,clip]{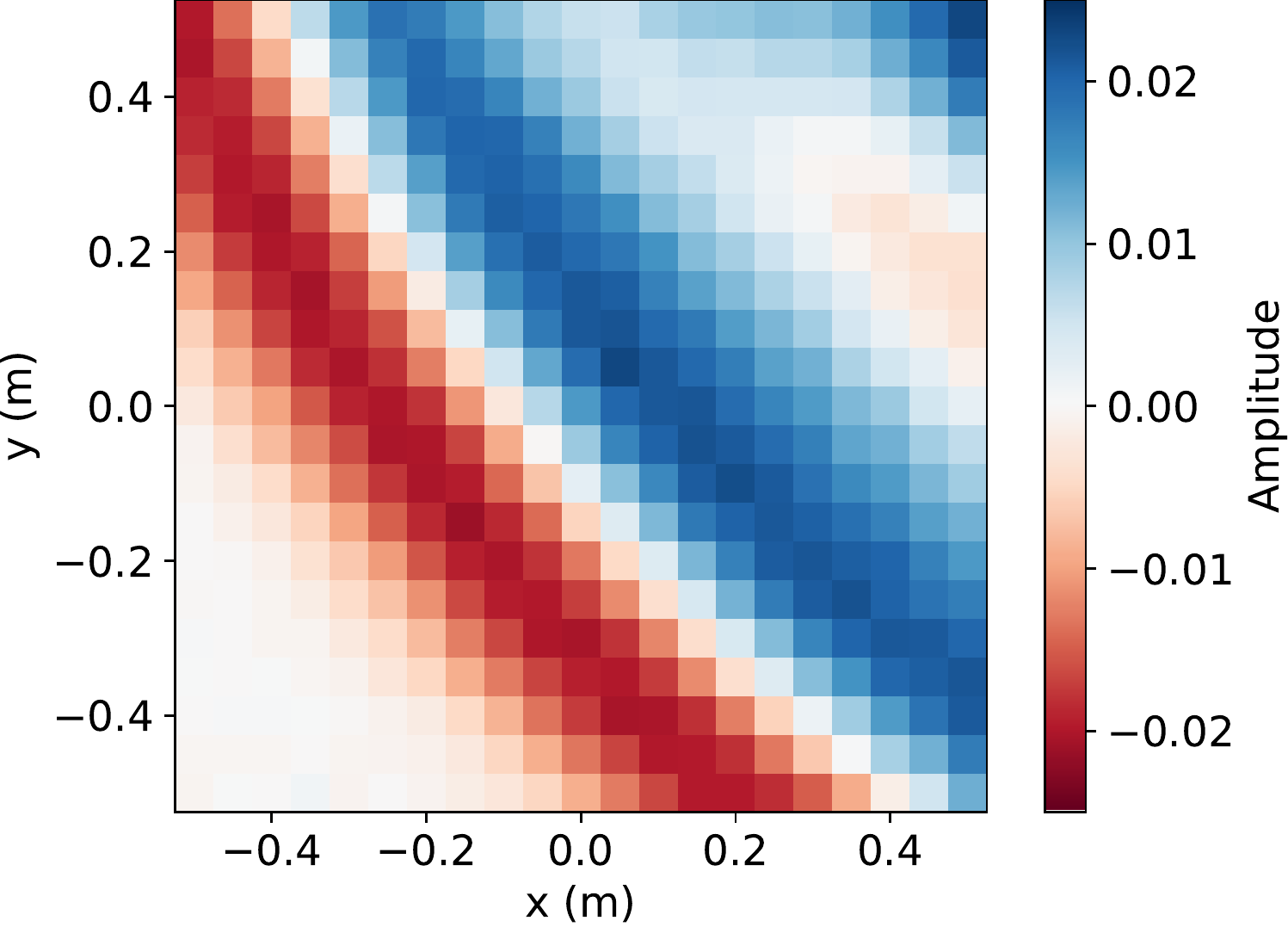}}
  \subfloat[$t=8600~\mathrm{sample}$]{\includegraphics[width=1.7in,clip]{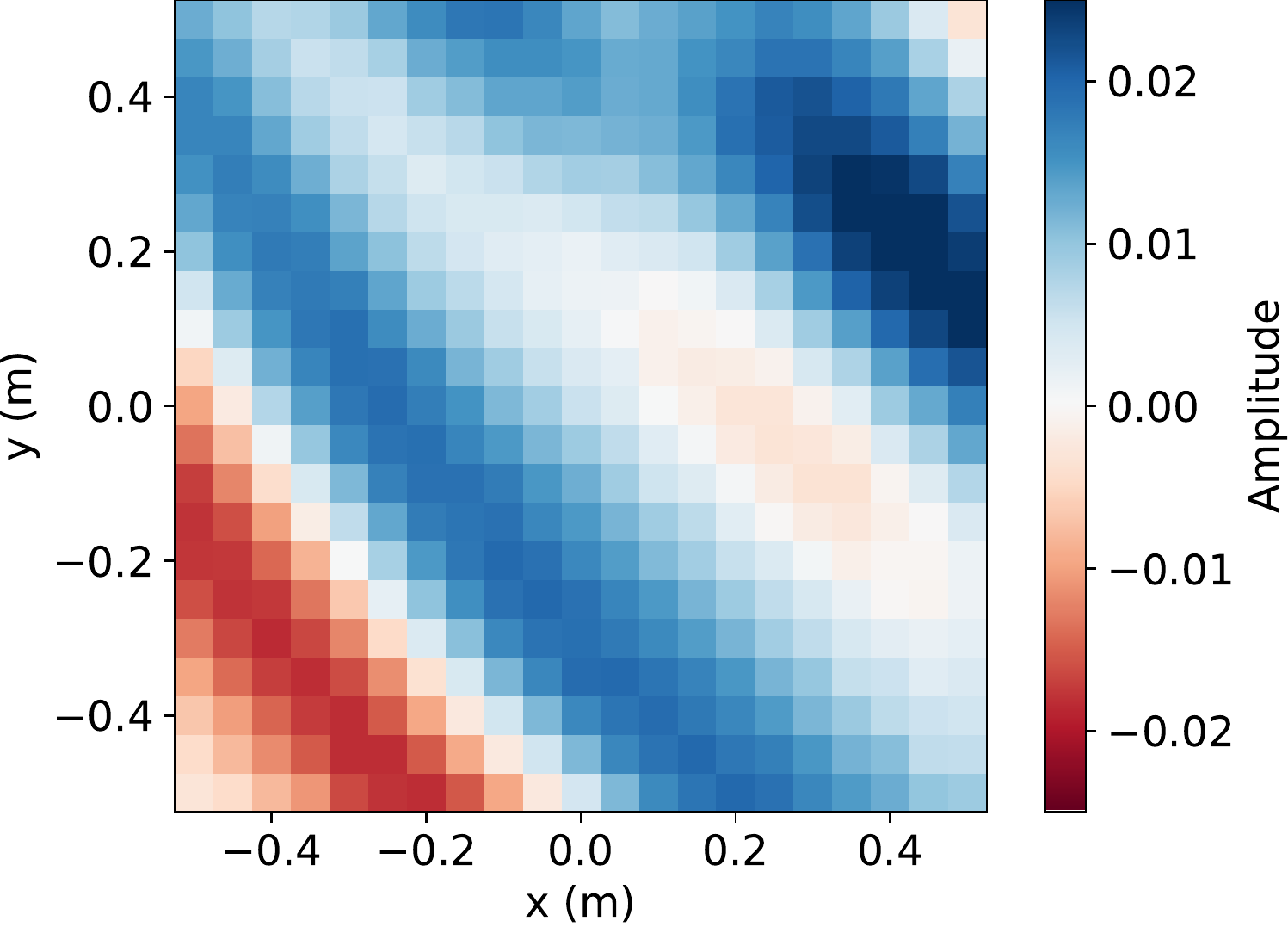}}
  \caption{Visualization of instantaneous pressure distributions using low-pass-filtered IRs of \textbf{S1-M3969}.}
  \label{fig:dist_xy}
%\vspace{-5pt}
\end{figure}
\begin{figure}[!t]
  \centering
  \includegraphics[width=3.2in,clip]{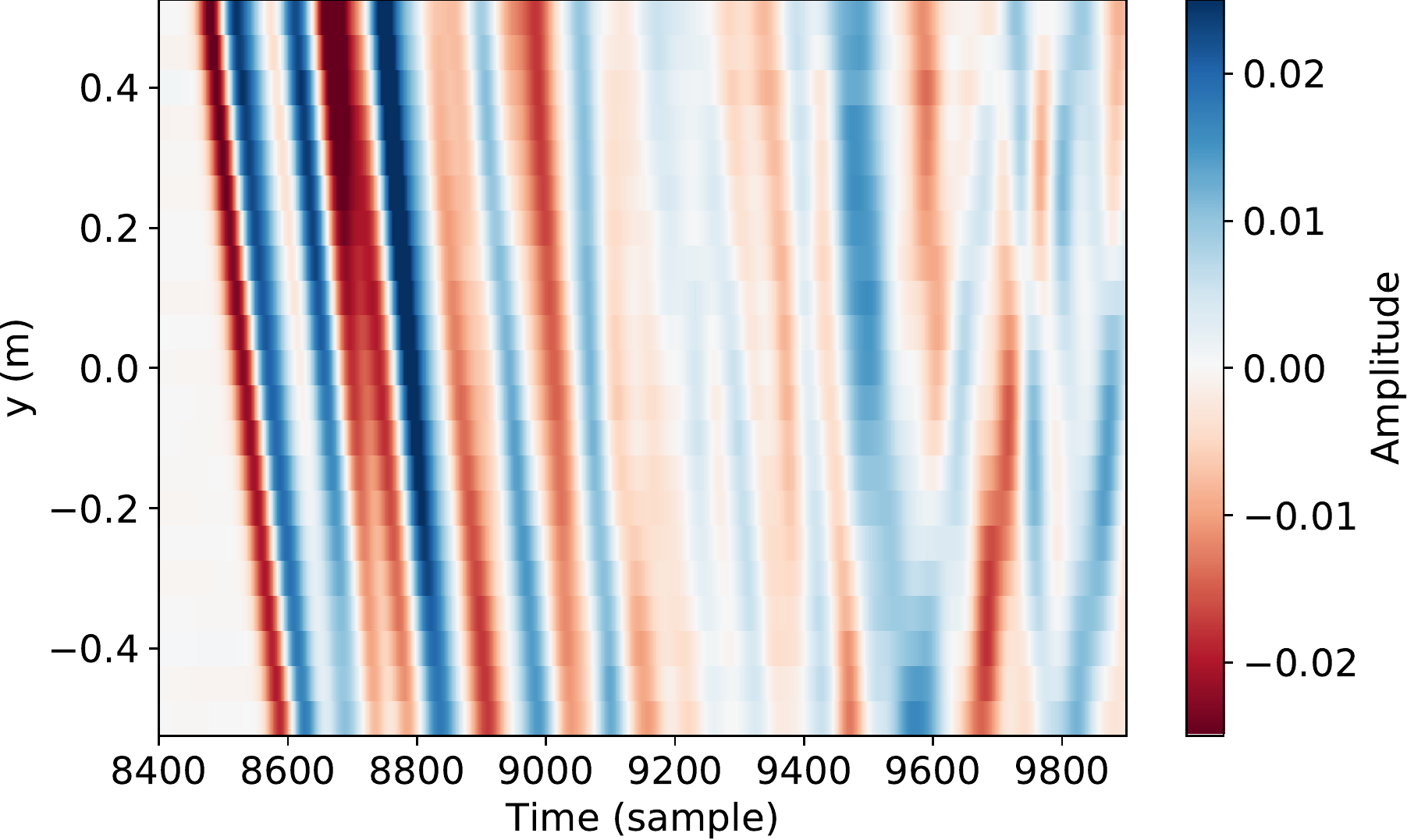}
  \caption{IRs with respect to time along $y$-axis at $(x,z)=(0,0)~\mathrm{m}$ of \textbf{S1-M3969}.}
  \label{fig:dist_yt}
%\vspace{-10pt}
\end{figure}

\begin{figure*}[!t]
  \centering
 \subfloat[True]{\includegraphics[width=2.2in,clip]{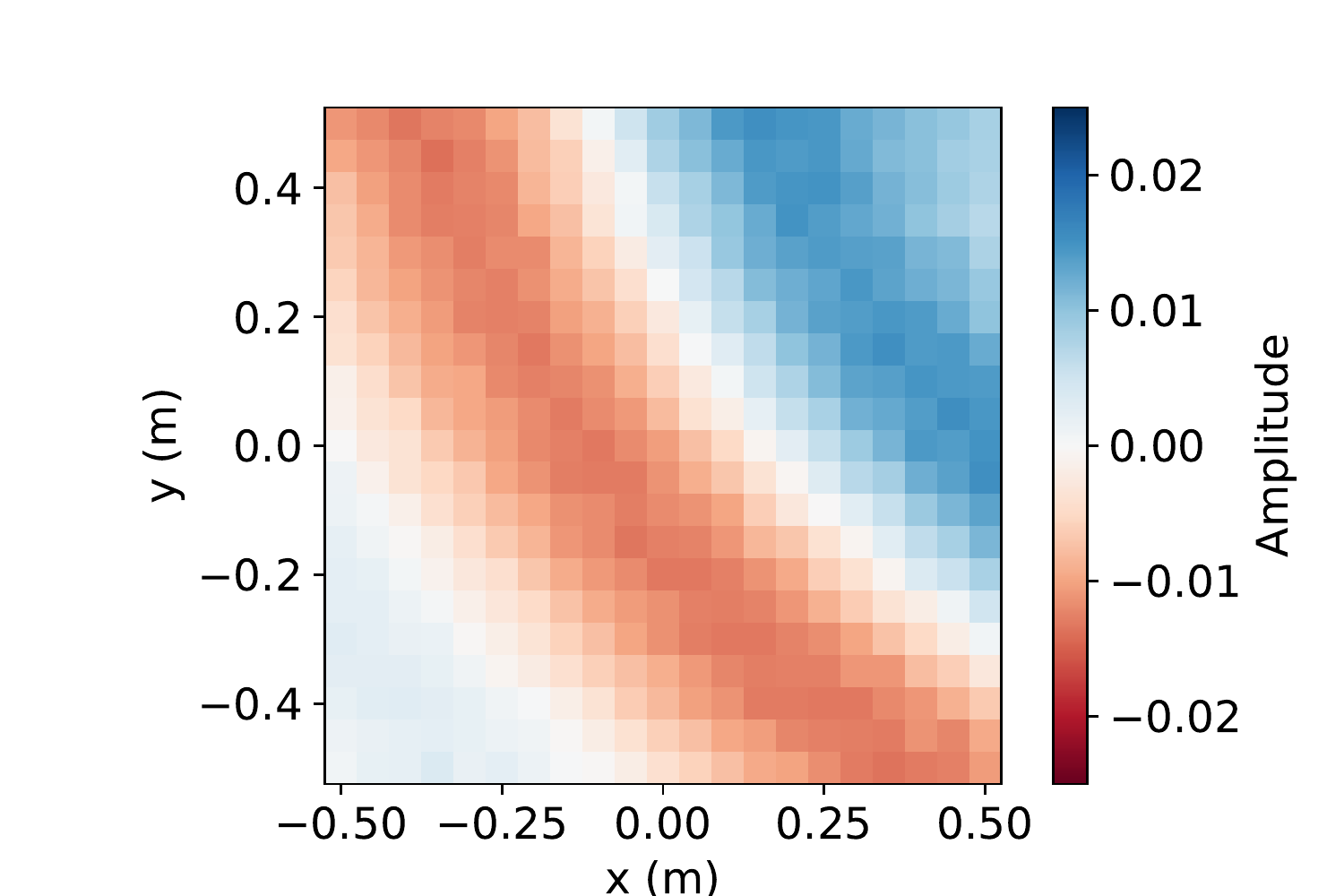}}
  \subfloat[Estimated]{\includegraphics[width=2.2in,clip]{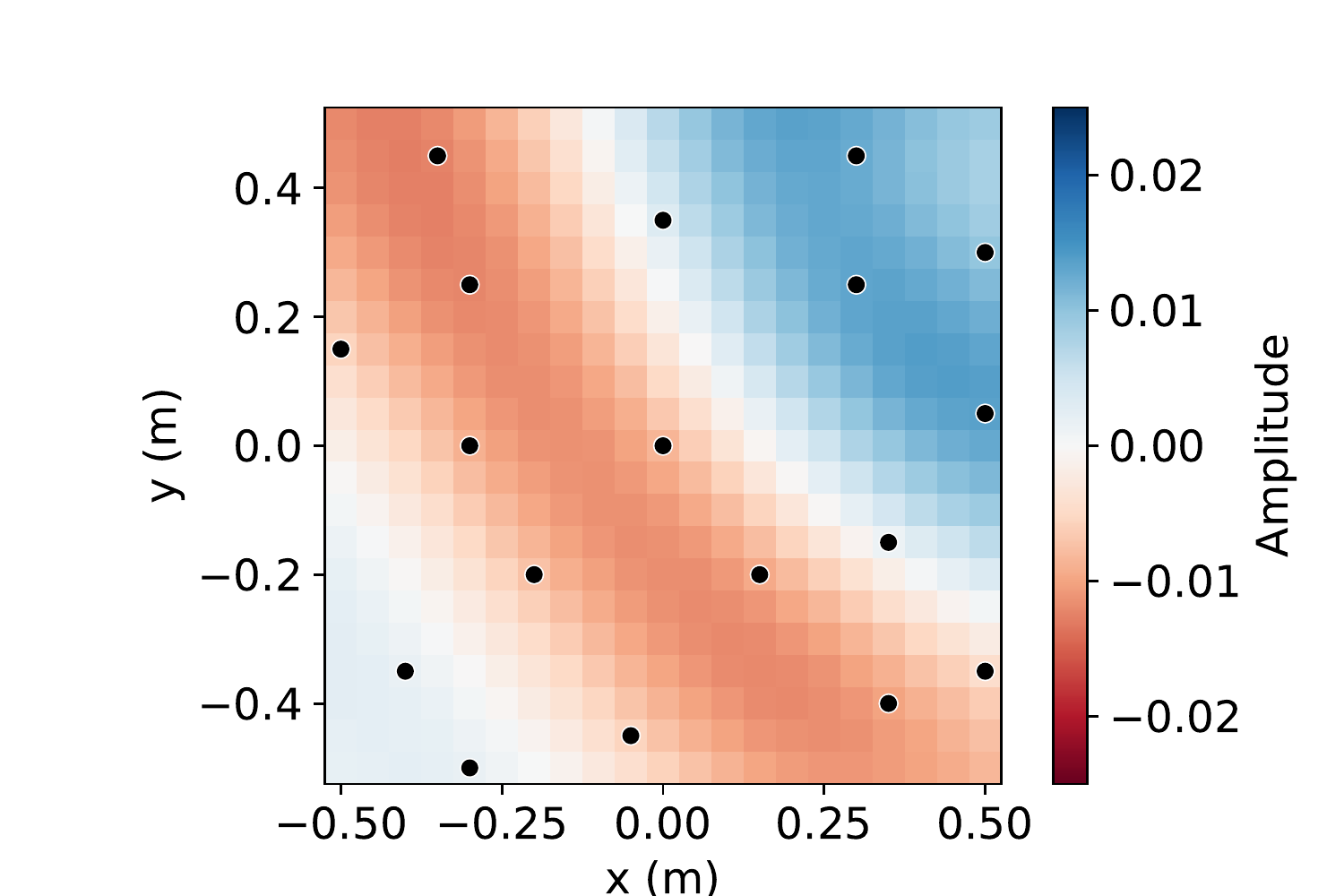}}
  \subfloat[Error]{\includegraphics[width=2.2in,clip]{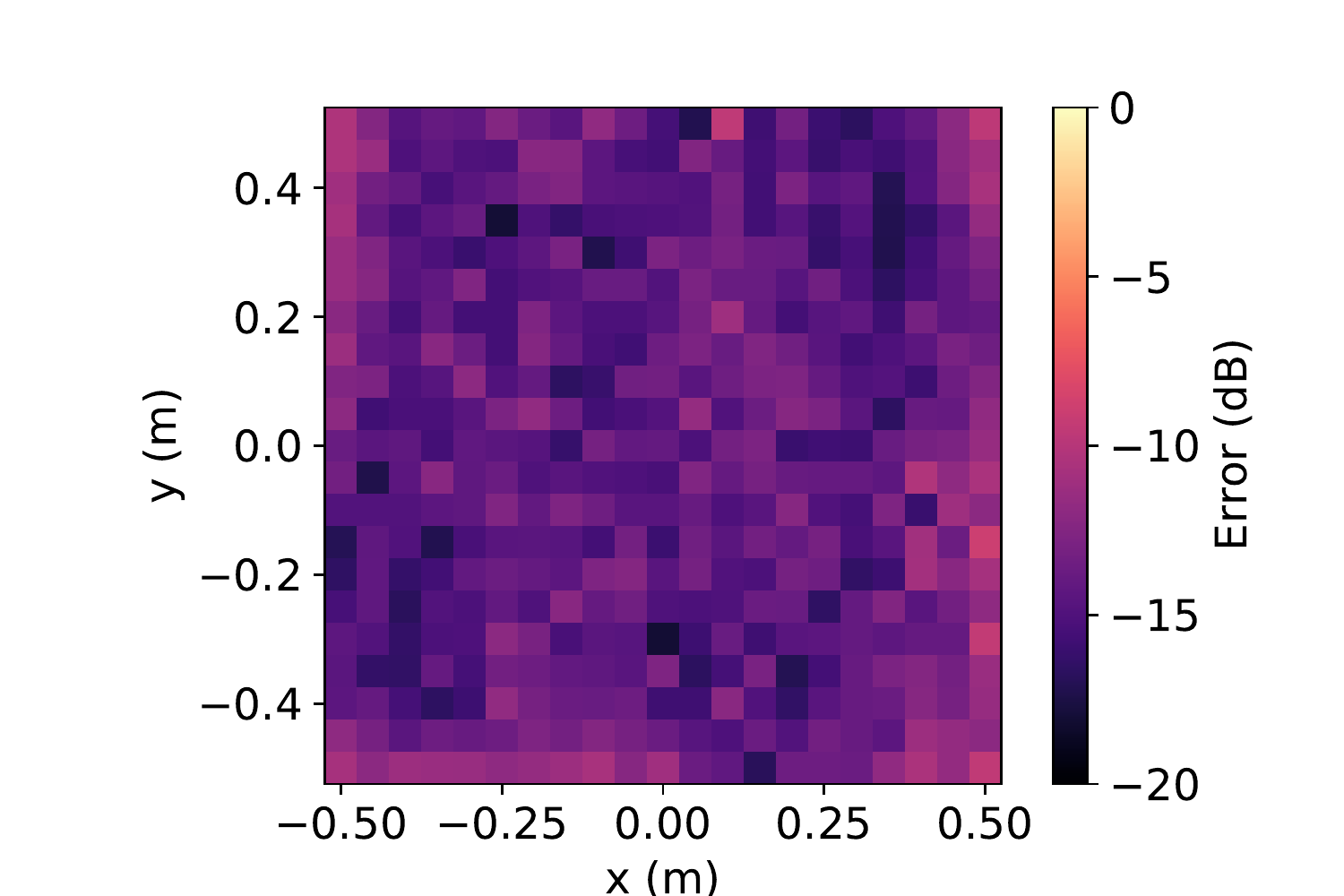}}
  \caption{Estimated pressure distribution at $t=0.089~\mathrm{s}$ and time-averaged square error distribution. Black dots in (b) indicate microphone positions. MSE was $-13.7~\mathrm{dB}$.}
  \label{fig:dist_reconst}
\vspace{-5pt}
%\end{figure*}
%\begin{figure*}[!t]
  \centering
 \subfloat[True]{\includegraphics[width=2.2in,clip]{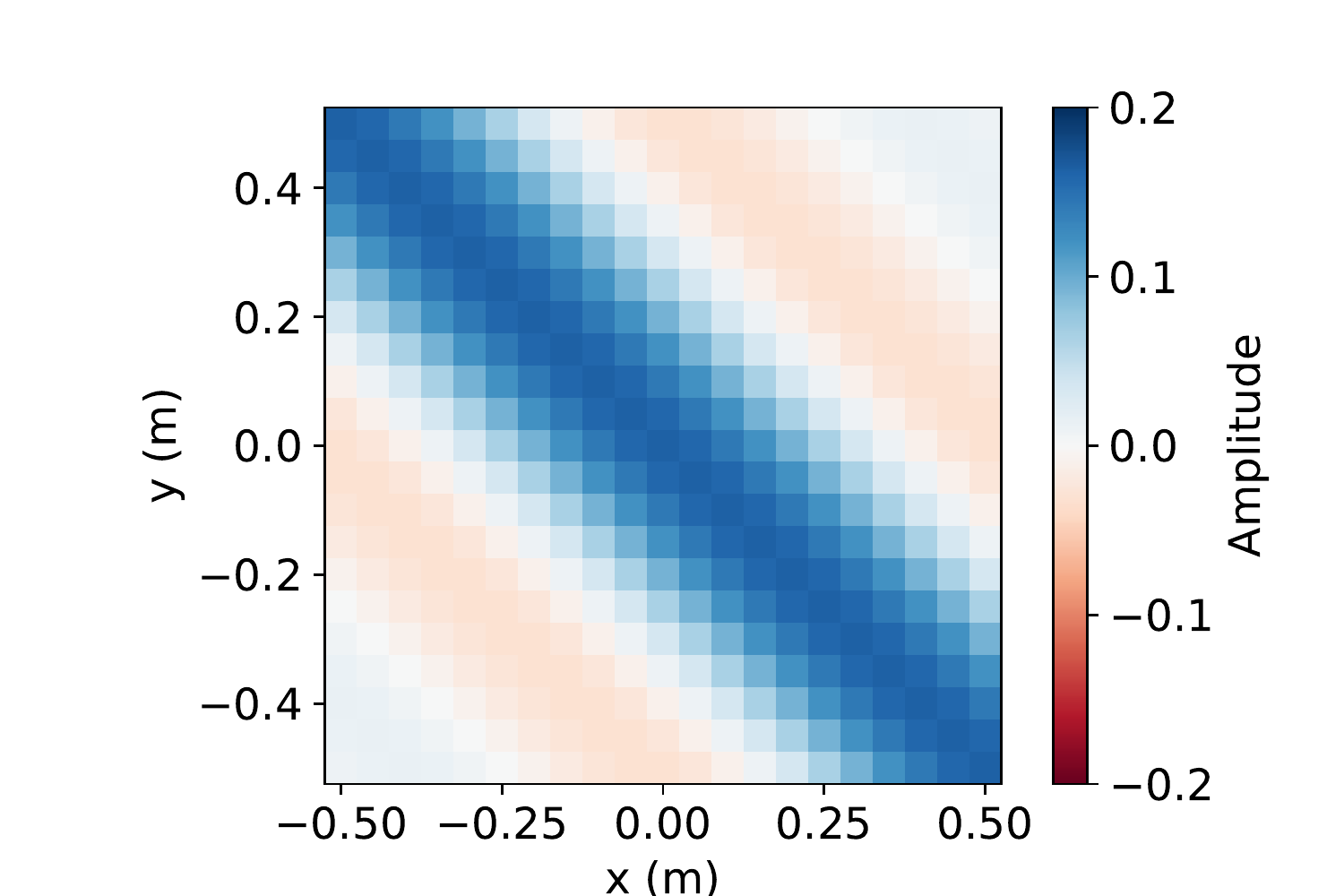}}
  \subfloat[Synthesized]{\includegraphics[width=2.2in,clip]{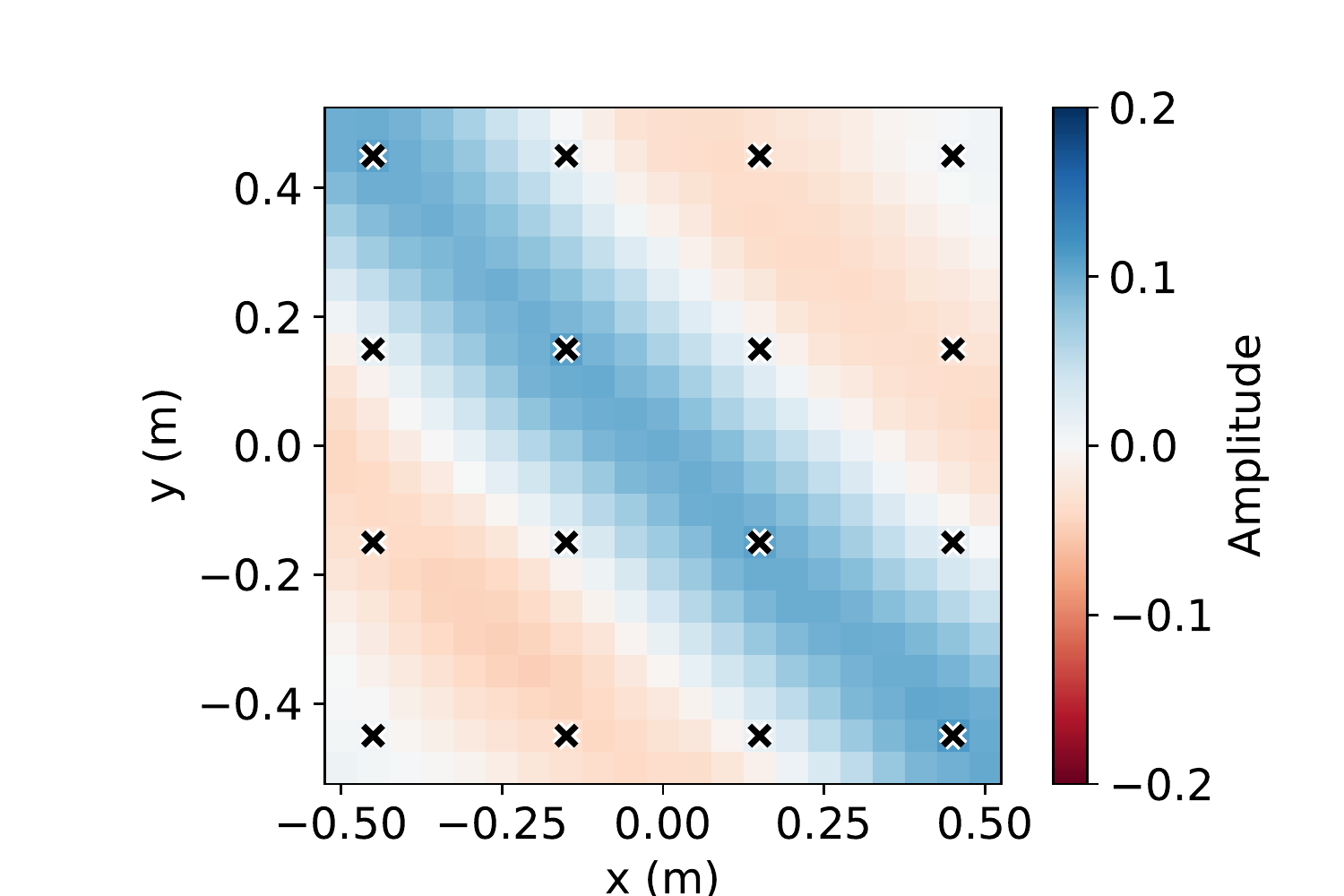}}
  \subfloat[Error]{\includegraphics[width=2.2in,clip]{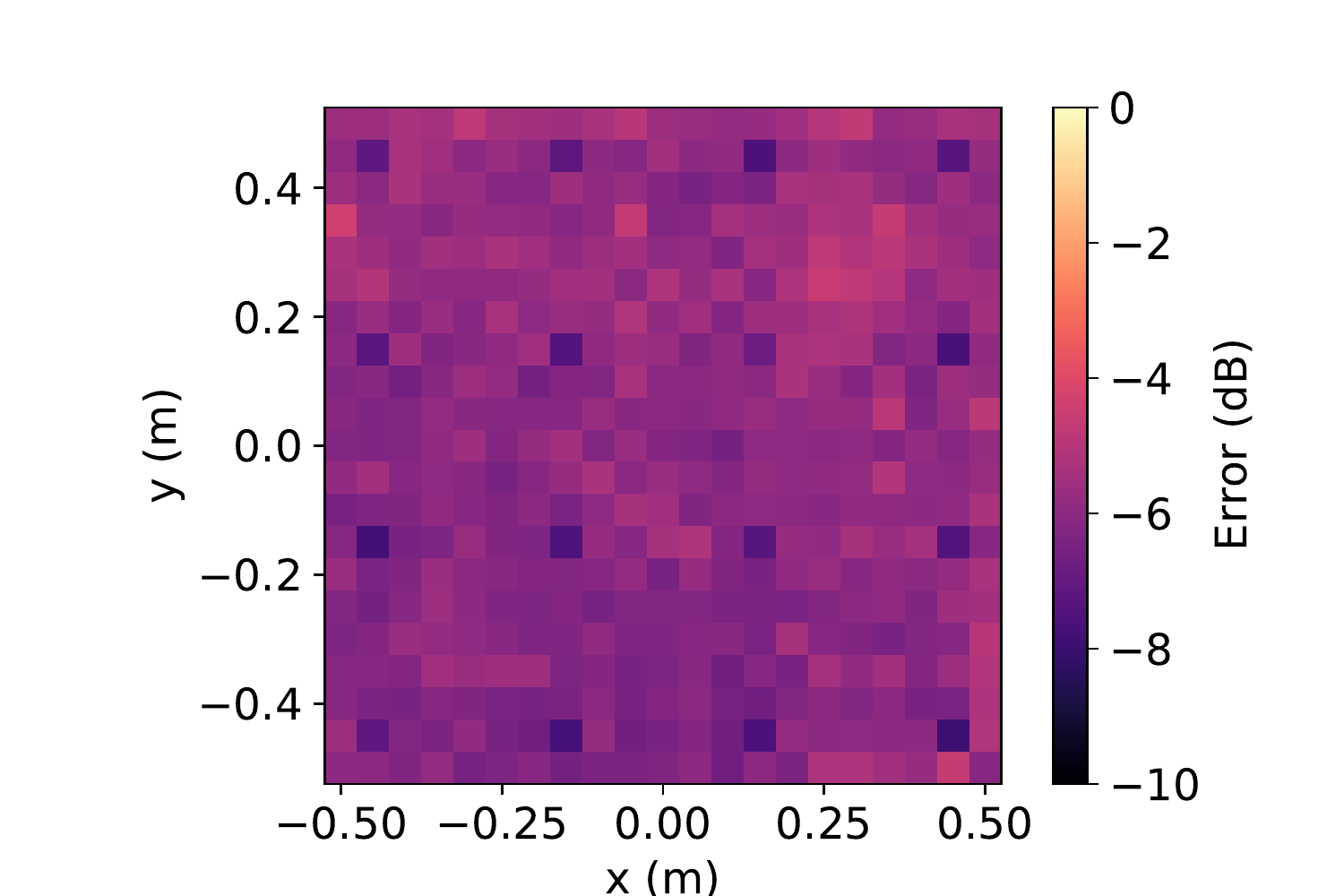}}
  \caption{Synthesized pressure distribution at $t=0.51~\mathrm{s}$ and time-averaged square error distribution. The black crosses in (b) indicate the positions of control points. The SDR was $3.85~\mathrm{dB}$.}
  \label{fig:dist_control}
\vspace{-8pt}
\end{figure*}

\subsection{Sound field reconstruction}

The problem of estimating and interpolating a sound field from a discrete set of microphone measurements is referred to as sound field reconstruction. We demonstrate experimental results of the sound field reconstruction using \textbf{S1-M3969}. 

The sound field reconstruction problem for a source-free region is generally solved by decomposing the measurements into element solutions of the Helmholtz equation using, for example, plane-wave and harmonic functions, which is a theoretical foundation of Fourier acoustics~\cite{Williams:FourierAcoust}. It is also possible to apply the equivalent source method~\cite{Johnson:JASA_J_1998}, which is based on the representation obtained using Green's function at a fictitious boundary enclosing the target region. Most of the current methods are based on the sound field decomposition into finite-dimensional basis functions. We here apply the spherical harmonic analysis of infinite order~\cite{Ueno:IEEE_SPL2018,Ueno:IEEE_J_SP_2021} to the estimation of the pressure distribution. When using omnidirectional microphones, this estimation method is equivalent to the kernel interpolation method using the zero-th-order spherical Bessel function as the kernel function~\cite{Ueno:IWAENC2018}. 

The target estimation region was set on the plane at the height of $z=0$. We set 18 microphones on this plane for pressure measurements, whose positions were chosen using the method proposed in~\cite{Nishida:EUSIPCO2020}. The source signal was a low-pass-filtered pulse signal, where the cuttoff frequency was $500~\mathrm{Hz}$, and the sampling rate was downsampled to $8000~\mathrm{Hz}$. The interpolation filter was designed in the time domain, and its length was $1025~\mathrm{samples}$.

Fig.~\ref{fig:dist_reconst} shows the estimated pressure distribution at $t=0.089~\mathrm{s}$ and the time-averaged square error distribution. The black dots in Fig.~\ref{fig:dist_reconst}(b) indicate the microphone positions. The mean square error (MSE) was $-13.7~\mathrm{dB}$. 

\subsection{Sound field control}

The sound field control is used for synthesizing a desired sound field inside a target region using multiple loudspeakers, which is applicable to high-fidelity spatial audio, personal audio created using multiple sound zones, and spatial active noise control. We synthesized a planewave field inside a target region using \textbf{S32-M441}. 

Methods of synthesizing a desired sound field based on integral equations analytically derived from the Helmholtz equation are basically applicable when the loudspeaker geometry is simple. On the other hand, a flexible loudspeaker geometry can be used in methods based on numerical optimization to minimize the error between synthesized and desired sound fields inside the target region. We here apply the pressure matching method~\cite{Kirkeby:JASA_J_1993}, which is widely used in practice because of its simple implementation. 

We set the target region as the measurement region, i.e., the square region with dimensions of $1.0~\mathrm{m} \times 1.0~\mathrm{m}$ at $z=0$. The desired sound field was a single plane wave whose arrival direction was $(\theta, \phi)=(\pi/2, \pi/4)$. 16 control points were set on the plane at intervals of $0.3~\mathrm{m}$. Again, the sampling rate was downsampled to $8000~\mathrm{Hz}$. The source signal was a low-pass-filtered pulse signal, where the cutoff frequency was $700~\mathrm{Hz}$. The control filter was designed in the time domain, and its length was $8192~\mathrm{samples}$.

Fig.~\ref{fig:dist_control} shows the synthesized pressure distribution at $t=0.51~\mathrm{s}$ and the time-averaged square error distribution. The black crosses in Fig.~\ref{fig:dist_control}(b) indicate the positions of the control points. The signal-to-distortion ratio (SDR) was $3.85~\mathrm{dB}$. The details on this sample application are described in~\cite{Koyama:I3DA2021}.

\section{Conclusion}
\label{sec:conclusion}

We introduced a new IR dataset called ``MeshRIR'' designed for the evaluation of sound field analysis and synthesis methods. An important feature of this dataset is that IR measurement points are obtained by finely discretizing a spatial region. Therefore, by choosing several points as microphone measurements, we can use the other points for validation. Several codes for sample applications are also provided. We plan to further extend the dataset to other source and microphone configurations and room conditions in future versions.

\section{ACKNOWLEDGMENT}
\label{sec:ack}

This work was supported by JST PRESTO Grant Number JPMJPR18J4.

%\newpage

% -------------------------------------------------------------------------
% Either list references using the bibliography style file IEEEtran.bst
\bibliographystyle{IEEEtran}
\bibliography{str_def_abrv,koyama_en,refs}
%
% or list them by yourself
% \begin{thebibliography}{9}
% 
% \bibitem{waspaa21web}
%   \url{http://www.waspaa.com}.
%
% \bibitem{IEEEPDFSpec}
%   {PDF} specification for {IEEE} {X}plore$^{\textregistered}$,
%   \url{http://www.ieee.org/portal/cms_docs/pubs/confstandards/pdfs/IEEE-PDF-SpecV401.pdf}.
%
% \bibitem{PDFOpenSourceTools}
%   Creating high resolution {PDF} files for book production with 
%   open source tools, 
%   \url{http://www.grassbook.org/neteler/highres_pdf.html}.
%
% \bibitem{eWilliams1999}
% E. Williams, \emph{Fourier Acoustics: Sound Radiation and Nearfield Acoustic
%   Holography}. London, UK: Academic Press, 1999.
% 
% \bibitem{ieeecopyright}
%   \url{http://www.ieee.org/web/publications/rights/copyrightmain.html}.
%
% \bibitem{cJones2003}
% C. Jones, A. Smith, and E. Roberts, ``A sample paper in conference
%   proceedings,'' in \emph{Proc. IEEE ICASSP}, vol. II, 2003, pp. 803--806.
% 
% \bibitem{aSmith2000}
% A. Smith, C. Jones, and E. Roberts, ``A sample paper in journals,'' 
%   \emph{IEEE Trans. Signal Process.}, vol. 62, pp. 291--294, Jan. 2000.
% 
% \end{thebibliography}

\end{sloppy}
\end{document}